\documentclass[fleqn,10pt,twoside]{article}
\usepackage{espcrc2}

\usepackage{graphicx}
\usepackage[figuresright]{rotating}

\newcommand{\AmS}{{\protect\the\textfont2
  A\kern-.1667em\lower.5ex\hbox{M}\kern-.125emS}}

\title{Charged hadron transverse momentum distributions in Au+Au collisions \\
at $\sqrt{s_{_{NN}}} =$ 200 GeV}

\author{Gerrit van Nieuwenhuizen for the PHOBOS Collaboration \\*[0.3cm]
\begin{small}
B.B.Back$^1$,
M.D.Baker$^2$,
D.S.Barton$^2$,
R.R.Betts$^6$,
M.Ballintijn$^4$,
A.A.Bickley$^7$,
R.Bindel$^7$,
A.Budzanowski$^3$,
W.Busza$^4$,
A.Carroll$^2$,
M.P.Decowski$^4$,
E.Garc\'{\i}a$^6$,
N.George$^{1,2}$,
K.Gulbrandsen$^4$,
S.Gushue$^2$,
C.Halliwell$^6$,
J.Hamblen$^8$,
G.A.Heintzelman$^2$,
C.Henderson$^4$,
D.J.Hofman$^6$,
R.S.Hollis$^6$,
R.Ho\l y\'{n}ski$^3$,
B.Holzman$^2$,
A.Iordanova$^6$,
E.Johnson$^8$,
J.L.Kane$^4$,
J.Katzy$^{4,6}$,
N.Khan$^8$,
W.Kucewicz$^6$,
P.Kulinich$^4$,
C.M.Kuo$^5$,
W.T.Lin$^5$,
S.Manly$^8$,
D.McLeod$^6$,
J.Micha\l owski$^3$,
A.C.Mignerey$^7$,
R.Nouicer$^6$,
A.Olszewski$^3$,
R.Pak$^2$,
I.C.Park$^8$,
H.Pernegger$^4$,
C.Reed$^4$,
L.P.Remsberg$^2$,
M.Reuter$^6$,
C.Roland$^4$,
G.Roland$^4$,
L.Rosenberg$^4$,
J.Sagerer$^6$,
P.Sarin$^4$,
P.Sawicki$^3$,
W.Skulski$^8$,
S.G.Steadman$^4$,
P.Steinberg$^2$,
G.S.F.Stephans$^4$,
M.Stodulski$^3$,
A.Sukhanov$^2$,
J.-L.Tang$^5$,
R.Teng$^8$,
A.Trzupek$^3$,
C.Vale$^4$,
G.J.van~Nieuwenhuizen$^4$,
R.Verdier$^4$,
B.Wadsworth$^4$,
F.L.H.Wolfs$^8$,
B.Wosiek$^3$,
K.Wo\'{z}niak$^3$,
A.H.Wuosmaa$^1$,
B.Wys\l ouch$^4$\\
$^1$~Argonne National Laboratory, Argonne, IL 60439-4843, USA\\
$^2$~Brookhaven National Laboratory, Upton, NY 11973-5000, USA\\
$^3$~Institute of Nuclear Physics, Krak\'{o}w, Poland\\
$^4$~Massachusetts Institute of Technology, Cambridge, MA 02139-4307, USA\\
$^5$~National Central University, Chung-Li, Taiwan\\
$^6$~University of Illinois at Chicago, Chicago, IL 60607-7059, USA\\
$^7$~University of Maryland, College Park, MD 20742, USA\\
$^8$~University of Rochester, Rochester, NY 14627, USA\\
\end{small}
}

\date{\today}
       
\begin{document}

\begin{abstract}
We present transverse momentum distributions of charged hadrons 
produced in Au+Au collisions at $\sqrt{s_{_{NN}}} =$ 200 GeV. The evolution 
of the spectra for transverse momenta $p_T$ from 0.25 to 5~GeV/C 
is studied as a function of collision centrality.
We find a significant change of the spectral shape between proton-antiproton 
and peripheral Au+Au collisions. When comparing peripheral to central Au+Au 
collisions, we find that the yields at the highest $p_T$ exhibit 
approximate scaling with the number of participating nucleons, rather than
scaling with the number of binary collisions.\\
\end{abstract}

\maketitle

\begin{figure}[htb]
\begin{minipage}[t]{77mm}
\includegraphics[width=7.5cm,height=8cm]{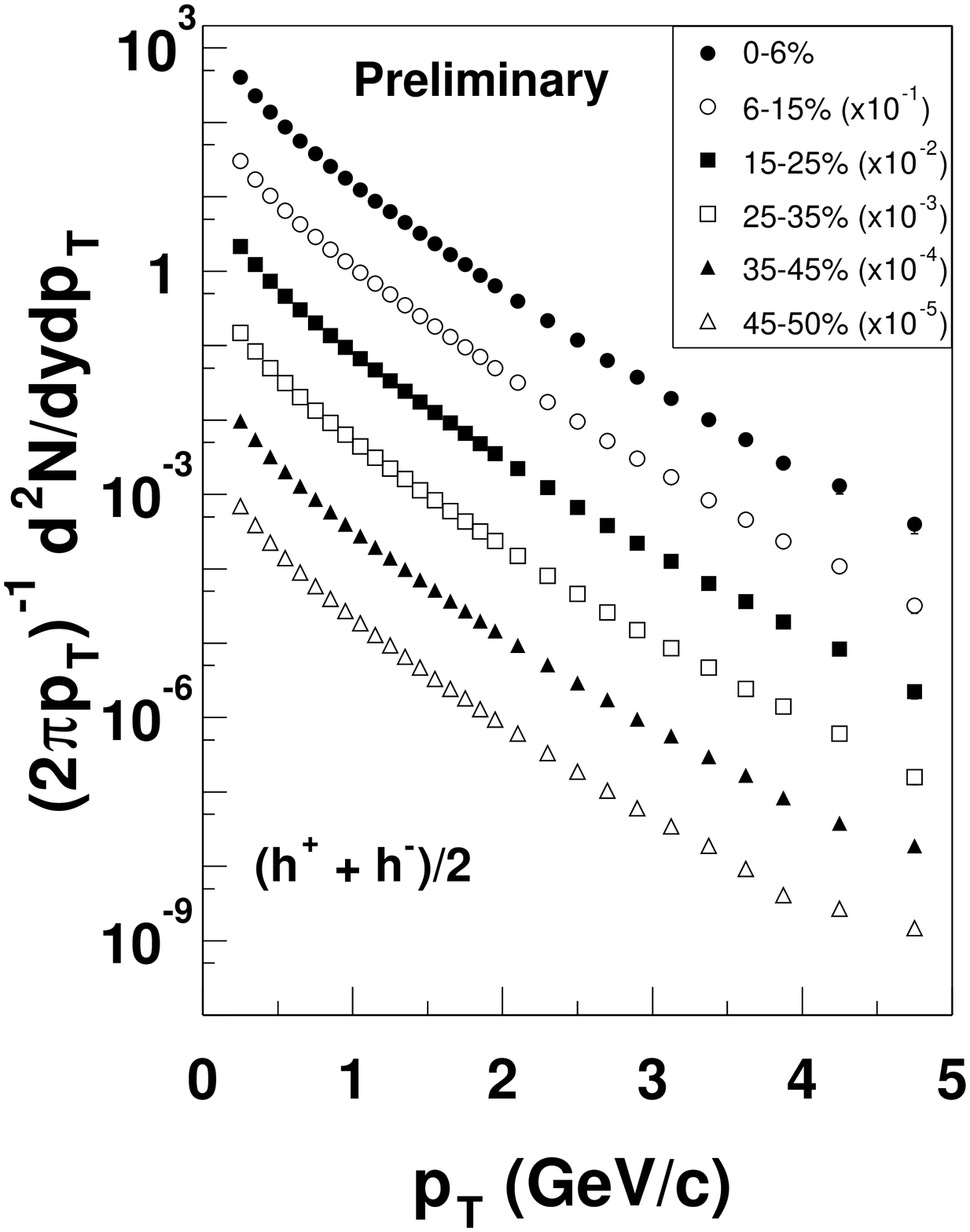}
\vspace{-1.0cm} 
\caption{ \label{ref_SpectraAllCent} 
Invariant yields for charged hadrons as a function of $p_T$ for 6 
centrality bins. For clarity, consecutive bins are scaled by factors of 10.}
\end{minipage}
\hspace{\fill}
\begin{minipage}[t]{77mm}
\includegraphics[width=7.5cm,height=8cm]{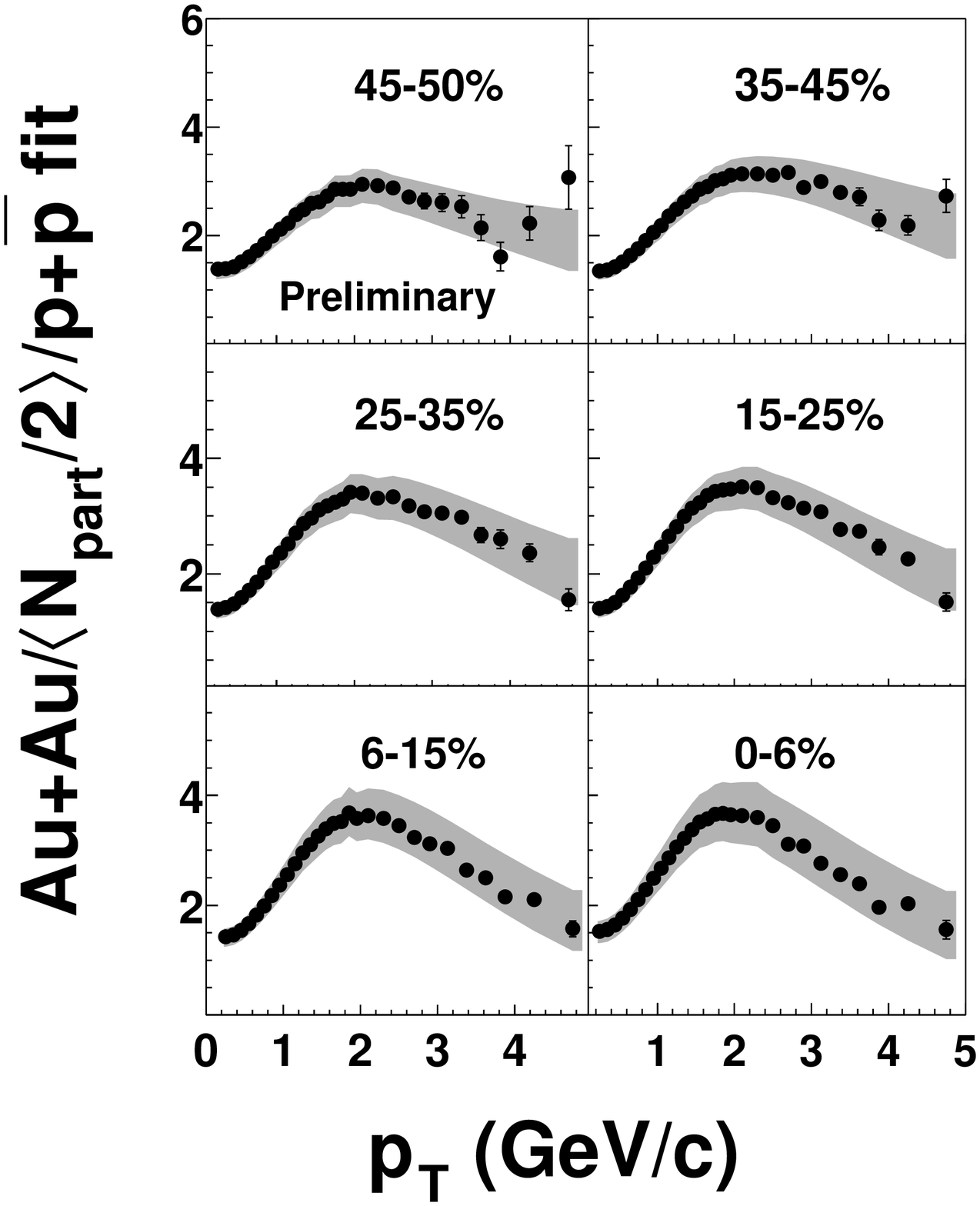}
\vspace{-1.0cm} 
\caption{ \label{ref_SpectraRatioPbarP}
Ratio of the yield of charged hadrons in Au+Au collisions to a fit of proton-antiproton data 
scaled by $\langle N_{part} /2 \rangle$ as a function of $p_T$. 
}
\end{minipage}
\end{figure}

\begin{figure}[htb]
\begin{minipage}[t]{77mm}

\includegraphics[width=7.5cm,height=8cm]{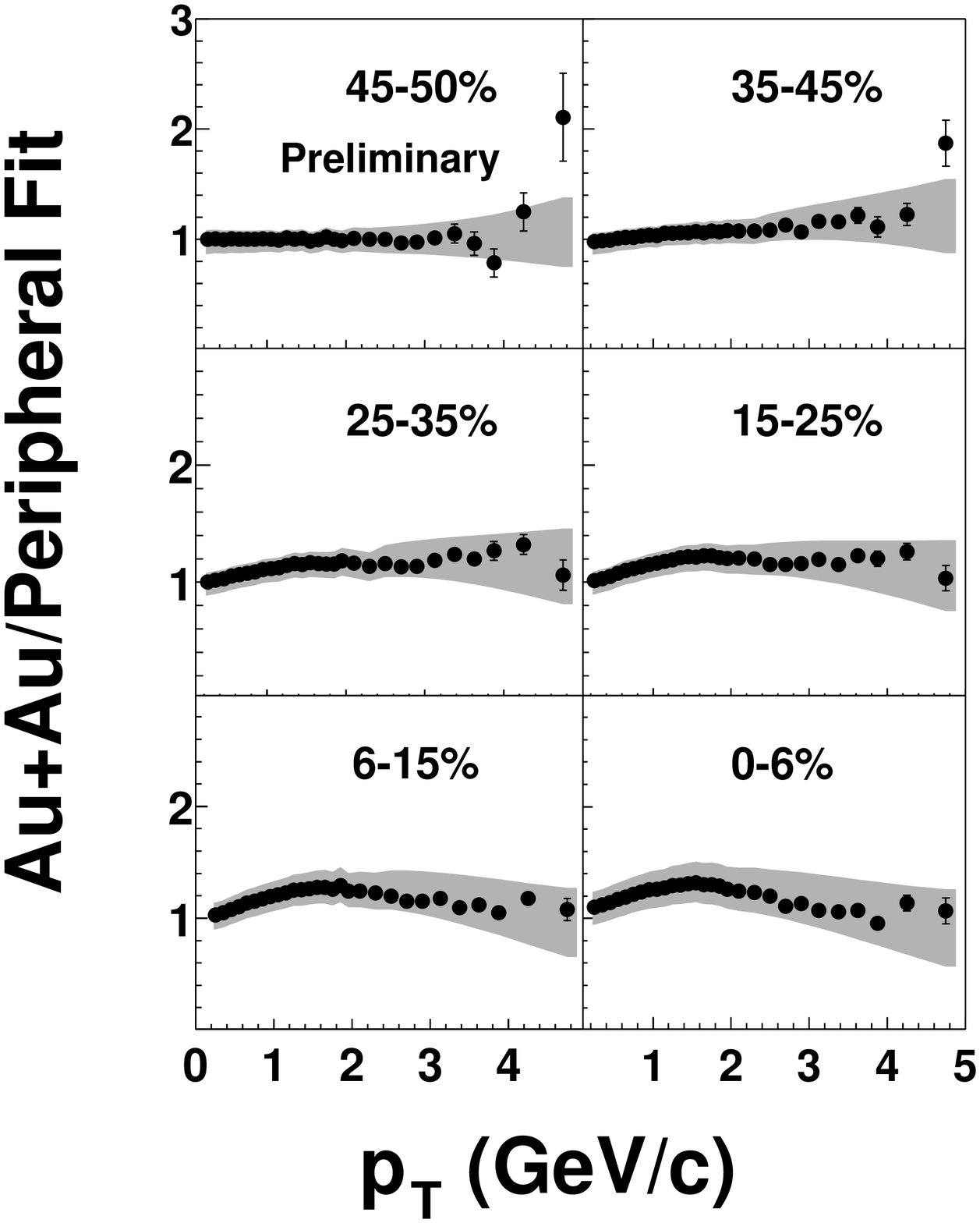}
\vspace{-1.0cm} 
\caption{
\label{ref_RatioPeripheral}
Charged hadron yield in Au+Au in six centrality bins, divided by
a fit to the most peripheral bin and scaled by
$\langle N_{part} /2 \rangle$.}
\end{minipage}
\hspace{\fill}
\begin{minipage}[t]{77mm}
\includegraphics[width=7.5cm,height=8cm]{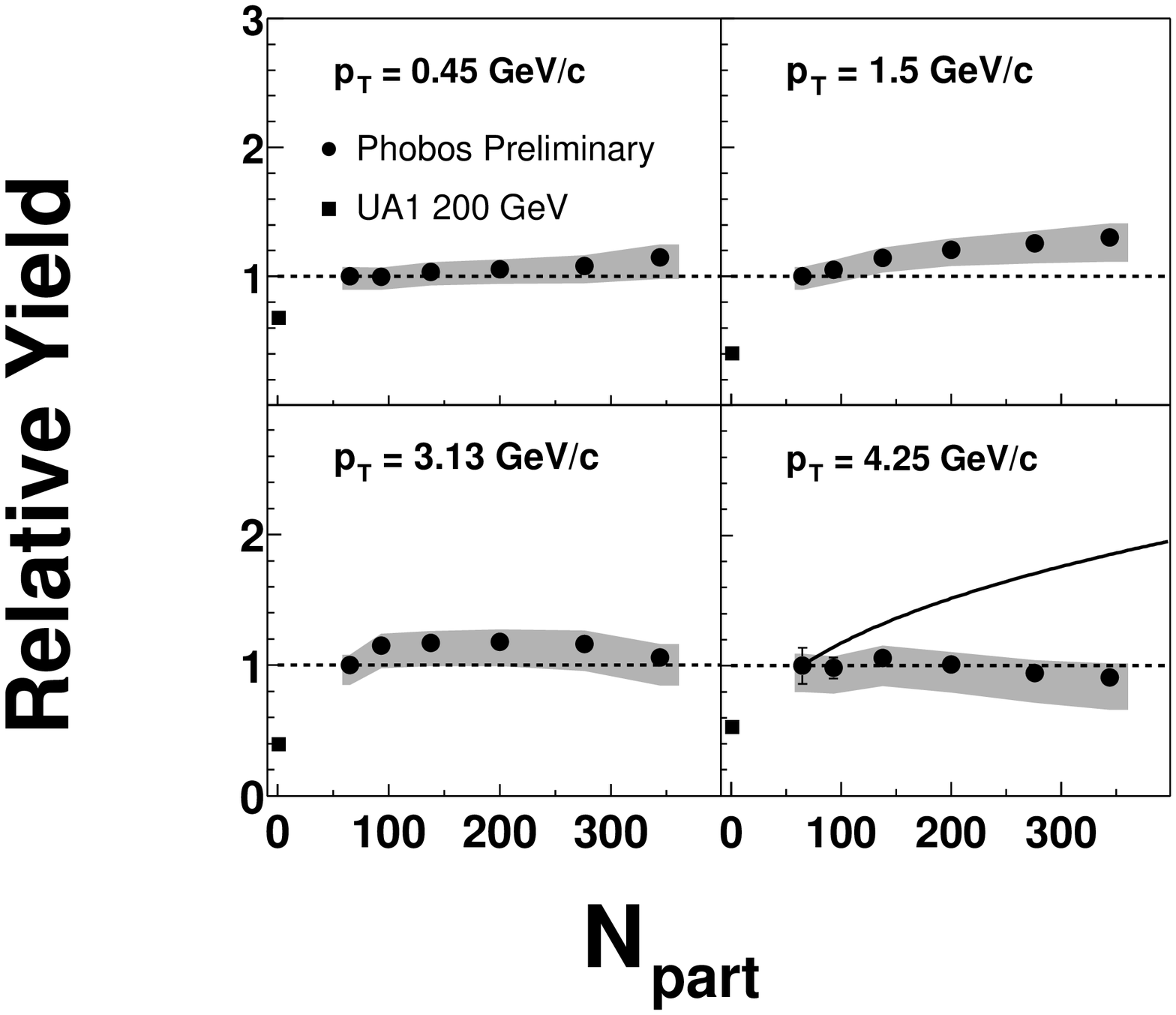}
 \vspace{-1.0cm} 
\caption{
\label{ref_YieldCent}
Charged hadron yield in Au+Au, normalized to the yield 
of the most peripheral bin, as  a function of $N_{part}$  
at different $p_T$. The 
solid line in the right bottom panel shows the expectation from $\langle N_{coll} \rangle$
scaling.
}
\end{minipage}
\end{figure}
%

%
%

In this paper, the yield of charged hadrons produced in collisions of gold nuclei at 
an energy of $\sqrt{s_{_{NN}}} = 200$~GeV is presented as a function of 
collision centrality and transverse momentum $p_T$.
The data were taken with the PHOBOS detector \cite{phobos1,phobos2}
during the second run of the Relativistic Heavy-Ion Collider (RHIC)
at the Brookhaven National Laboratory. 
In the theoretical analysis of particle production in hadronic and nuclear collisions,
a distinction is often made between the relative contributions of ``hard'' parton-parton
scattering processes and ``soft'' processes. Hard processes
are expected to contribute an increasingly larger fraction of particle production
with increasing collision energy and at higher $p_T$ of the produced particles.

Collisions of heavy nuclei offer ideal conditions to test our understanding 
of this picture, as ``hard'' processes are expected to scale with 
the number of binary nucleon-nucleon collisions $N_{coll}$, whereas ``soft'' 
particle production is expected to exhibit scaling with the number of participating 
nucleons $N_{part}$. In the Glauber picture of nuclear collisions, $N_{coll}$ approximately scales as 
$(N_{part})^{4/3}$. For central collisions of  Au nuclei, one therefore obtains 
an increase in the ratio of $N_{coll}/(N_{part}/2)$ by a factor of six, relative 
to proton-proton collisions. 

%
%
In Fig.~\ref{ref_SpectraAllCent} we present the invariant cross-section
of charged hadrons
as a function of transverse momentum for particles with a rapidity
of $0.2 < y_\pi < 1.4$ assuming the pion mass for calculating $y_\pi$.
The cross-sections were corrected for the geometrical
acceptance of the detector, the reconstruction efficiency, the distortion
due to binning and momentum resolution, ``ghost'' tracks, secondaries and
weak decays \cite{highpt,pbarp_200}.

%
%
Data are shown for 6 centrality bins, with $65 < \langle N_{part}\rangle < 344$ corresponding,
in the Glauber picture,
to $107<\langle N_{coll}\rangle<1050$.
The increase with centrality of the integrated yields, over
the studied centrality range, is consistent with the centrality
evolution of the mid-rapidity
particle density presented in \cite{phobos_cent_200}.

The top left panel in Fig.~\ref{ref_SpectraRatioPbarP} shows the ratio of the most peripheral bin
($\langle N_{part} \rangle = 65 \pm 4)$ divided by
$\langle N_{part}/2\rangle \times f(p_T)$, where $f(p_T)$ is a fit of the invariant 
cross-section in proton-antiproton collisions at the same energy \cite{ua1_pbarp}.
This comparison shows that already in Au+Au collisions with an impact parameter
of about 10~fm, the spectral shape is dramatically modified relative to that
in $p+\bar{p}$ collisions.
It is worth noting
that the ratio 
$\langle N_{coll}\rangle{/}\langle N_{part} / 2\rangle$ 
increases
by a factor of about three from $p+\bar{p}$ to peripheral Au+Au collisions.
The other panels in Fig.~\ref{ref_SpectraRatioPbarP} show the evolution of
same ratio upto the most central bin  ($\langle N_{part} \rangle = 344 \pm 12)$.
We observe only a moderate change 
in spectral shape, relative to $p+\bar{p}$, when 
going from peripheral to central events.

Fig.~\ref{ref_RatioPeripheral} shows the evolution of the spectra 
with centrality.
The spectra, normalized by $\langle N_{part}/2\rangle$, in the six centrality
bins have been divided by a fit to the normalized spectrum measured in the 
most peripheral bin.
The centrality range in the Au+Au collisions
covers a change in 
$\langle N_{coll}\rangle{/}\langle N_{part} / 2\rangle$ 
by
about a factor of two. 
It is remarkable that the change in spectral shape over
this range of centralities is small compared to that between peripheral
Au+Au collisions 
and $p+\bar{p}$ collisions.
In particular at high $p_T$, above 3~GeV/c, the yields
scale to a good approximation as $\langle N_{part} / 2\rangle$ as a function of centrality.

Fig.~\ref{ref_YieldCent} shows the scaling behaviour of the charged hadron relative yields in
Au+Au and $p+\bar{p}$ collisions for different values of $p_T$.
The yields were determined by interpolating the data shown in 
Fig.~\ref{ref_SpectraAllCent} using power law fits.

%
%
 The observed trends can be contrasted with the expectation that 
particle production should be characterized by a change from  $\langle N_{part}\rangle$ 
scaling at low $p_T$ to $\langle N_{coll}\rangle$ scaling at high $p_T$.
No corresponding increase in particle 
production per participant at $p_T = 3$~GeV/c and above is observed. Rather,
the yields in this region scale approximately with the number of participating nucleons.
Upcoming studies of d+Au collisions at RHIC will provide further
insight into the modification of particle spectra in a nuclear environment.\\

This work was partially supported by US DoE grants DE-AC02-98CH10886,
DE-FG02-93ER-404802, DE-FC02-94ER40818, DE-FG02-94ER40865, DE-FG02-99ER41099, W-31-109-ENG-38.
NSF grants 9603486, 9722606 and 0072204. The Polish groups were partially supported by KBN grant 2-P03B-103203.
The NCU group was partially supported by NSC of Taiwan under 
contract NSC 89-2112-M-008-024.

\end{document}